\documentclass[pra,aps,twocolumn]{revtex4}

\usepackage{epsfig}
\usepackage{float}

\def\>{\rangle}

\def\F{\mathcal{F}}
\def\S{\mathcal{S}}
\def\E{\mathcal{E}}
\def\P{\mathcal{P}}
\def\ot{\otimes}

\def\rin{\rho_{\rm in}}
\def\ein{e_{\rm in}}
\def\psiin{|\psi_{\rm in}\rangle}
\def\cein{|e_{\rm in}\rangle}

\def\rout{\rho_{\rm out}}
\def\eout{e_{\rm out}}
\def\eoutid{e^{\rm ideal}_{\rm out}}

\def\ceout{|e_{\rm out}\rangle}

\def\sx{\sigma_{\rm x}}
\def\sy{\sigma_{\rm y}}
\def\sz{\sigma_{\rm z}}

\newcommand{\be}{\begin{equation}}
\newcommand{\ee}{\end{equation}}

\newcommand{\bea}{\begin{eqnarray}}
\newcommand{\eea}{\end{eqnarray}}

\newcommand{\eq}[1]{Eq.~(\ref{eq:#1})}

\begin{document}
\title{Simple proof of fault tolerance in the graph-state model}

\author{Panos Aliferis$^1$ and Debbie W. Leung$^{1,2}$}

\affiliation{$^1$Institute for Quantum Information, California Institute of Technology, Pasadena, CA 91125, USA}

\affiliation{$^2$Institute for Quantum Computing, University of Waterloo, Waterloo, ON N2L 3G1, Canada}

\date{March 08, 2006}

\begin{abstract}

We consider the problem of fault tolerance in the graph-state model of quantum computation. Using the notion of composable simulations, we provide a simple proof for the existence of an accuracy threshold for graph-state computation by invoking the threshold theorem derived for quantum circuit computation. Lower bounds for the threshold in the graph-state model are then obtained from known bounds in the circuit model under the same noise process.

\end{abstract}

\maketitle

\raggedbottom

\section{Introduction}

Alternative models for quantum computation based on projective measurements \cite{Raussen01, Nielsen01, Leung03} have recently attracted much attention. A common concept of these models is the simulation of individual quantum circuit operations and how simulations can be composed together \cite{Childs04}. More specifically, in these models a sequence of single- or two-qubit measurements is applied to a collection of fixed initial quantum states thereby in effect simulating unitary transformations on a smaller subspace of states. 

Although our results are applicable to a much larger class of measurement-based models, our analysis will focus on a variation of Raussendorf and Briegel's one-way quantum computer model \cite{Raussen01} where computation is performed by single-qubit projective measurements on 
some initial \emph{graph state} \cite{Raussen03}. Henceforth, we will refer to this model as the {\em graph-state model} and the computation realized in it as a {\em graph-state simulation}. 

The graph-state model offers a decomposition of a quantum algorithm in terms of alternative elementary primitives, as well as potential advantages in certain physical implementations. For example, suppose entangling gates can only be realized nondeterministically with flagged faults as, e.g., in optical quantum computation \cite{Knill00}.  Then, graph-state simulation offers much advantage since entangling gates are only used for the preparation of the graph state, which can be done independently from the main computation \cite{Nielsen04}.  

In any physical realization of quantum computation, unknown errors will always be present and they will have to be corrected using quantum error-correcting codes in a fault-tolerant manner. 
An important question is therefore under what conditions computation can be executed reliably in the graph-state model in the presence of physical noise. In the circuit model, fault-tolerant methods \cite{Shor96} are available for the reliable execution of any desired computation, if the noise is sufficiently weak. Now, if such a fault-tolerant circuit is {\em simulated} in the graph-state model with sufficiently weak noise, will the same desired computation be reliably executed? More specifically, this is answered by first analyzing noisy simulations of individual operations and then how the noisy simulations compose together. 

The first results on this problem were reported in the Ph.D.\ thesis of Raussendorf \cite{Raussen03b}. This work proved the existence of an accuracy threshold for cluster-state computation for various independent stochastic error models. More recently, Nielsen and Dawson \cite{Nielsen04b} obtained proofs for the existence of an accuracy threshold in the graph-state model that apply to more general error models (including errors due to nondeterministic gates) by reduction to a threshold theorem for local non-Markovian noise \cite{Terhal04}. In addition, they established a conceptual framework and two technical theorems that are of independent interest.

%
In this paper, we use the concept of composable simulations \cite{Childs04} and the threshold theorem derived in the circuit model \cite{Aharonov99comb, Knill96b, Kitaev97b, Terhal04, Aliferis05b} to obtain a simple proof for the existence of an accuracy threshold in the graph-state model. 
Furthermore, for any specific form of noise, our proof allows known lower bounds on the threshold in the circuit model to be translated to equivalent bounds in the graph-state model. We discuss in particular how the two lower bounds are related for commonly used fault-tolerant architectures based on self-dual CSS codes.


\section{Review of the graph-state model}

We begin by briefly introducing the graph-state model 
in the ideal noiseless case. Various recent interpretations of this model have been reported and reviewed recently \cite{Verstraete03}, \cite{Aliferis04}, \cite{Childs04}, \cite{Perdrix04b}, \cite{Nielsen05b}, \cite{Jozsa05}.
We follow the language in Ref.$\,$\cite{Childs04}, which explicitly uses the notion of composable simulations that forms the core of our subsequent analysis. Our discussion in this section is intended to introduce the basic notions and terminology that we will use later in our proof. 

In the circuit model, an arbitrary quantum computation can be decomposed into state preparation, measurements, and a universal set of gates. To show the universality of the graph-state model of quantum computation, it suffices to show that (i) each element for universality in the standard model can be simulated and (ii) the simulation can be composed to simulate the entire computation.  The approach is to first define an appropriate notion of simulation that is composable, followed by a complete recipe to composably simulate each element needed for universality.  


We first describe the notion of composable simulations. Let $\F$ be an operation (a superoperator, or completely positive trace-preserving map) to be simulated and $\S_\F$ be the associated operation that simulates $\F$. For simplicity, let $\F$ act on $n$ qubits. In the general case, $\F$ can have quantum and classical input and output of arbitrary dimensions, but this only requires extra notations and therefore will not be written out explicitly here.  For a $2n$-bit string $x$, let $\P_x$ be the superoperator corresponding to conjugation by 
the Pauli operator indexed by $x$.  
Our composable simulation $\S_\F$ takes two inputs, a classical $2n$-bit string $\ein$ and an $n$-qubit quantum state $\P_{\ein} (\rin)$, so that $\forall \rin$, $\forall \ein$, it acts as
\be
\label{eq:compdef}
	\S_\F ( \ein \otimes \P_{\ein} (\rin) ) 
	= \sum\limits_{\eout} p_{\eout} \eout \otimes (\P_{\eout} \! \circ \F) (\rin) \, ,
\ee
\noindent where $\eout$ is some new $2n$-bit string that appears with probability $p_{\eout}$.  (Throughout the paper, the symbol for a bit string such as $\ein$ also labels the corresponding density matrix.) To rephrase the above definition, for each specific classical output $\eout$, $\S_\F$ evolves the arbitrary state $\rin$ according to the intended operation $\F$ up to a new known succeeding Pauli operation $\P_{\eout}$, despite the $\P_{\ein}$ occurring to $\rin$ prior to the simulation. 

Note that $\eout$ is a function of $\ein$ and the measurement outcomes obtained in $\S_\F$, and this function depends on $\S_\F$. 
%
However, the statistics of $\eout$ has no consequence, because composable simulations work for \emph{all} measurement outcomes and for all $\ein$---all outcome histories lead to valid simulations, where an ``outcome history'' denotes the set of all measurement outcomes collected in a specific run of the simulation. %
%
As we will see next, this is important as it will allow us to compose simulations of individual operations to obtain a simulation of the combined operation.

Now, consider simulating a sequence of $l$ operations $\{\F_j\}$, and we will see that it can be done by composing the sequence of simulations $\{\S_{\F_j}\}$. By repeated applications of \eq{compdef}, $\forall \rin$, $\forall \ein$,
\begin{equation}
\begin{array}{c}
\label{sequence}
\S_{\F_l} \circ \cdots \circ \S_{\F_1} (\ein \ot \P_{\ein} (\rin) ) \hspace*{15ex} \\[1.2ex]
= \sum\limits_{\eout} p_{\eout } \eout \ot (\P_{\eout} \circ \F_l \circ \cdots \circ \F_1) (\rin) \, ,
\end{array}
\end{equation}
\noindent which states that, for all outcome histories, the entire sequence of operations $\{\F_j\}$ is simulated properly, up to a final overall $\P_{\eout}$ (which just redefines the final classical outcome of the computation).  

We will now describe how composable simulations are realized in the graph-state model. Let $\Gamma$ denote a graph with vertex set $V(\Gamma)$ and edge set $E(\Gamma)$. 
One way to specify and to create the graph state corresponding to $\Gamma$ is to start with the initial state $\bigotimes_{i\in V(\Gamma)} |+\>$ and then apply a controlled-phase ({\sc cphase}) gate to each pair of qubits in $E(\Gamma)$ (where {\sc cphase}$\,|ab\> = (-1)^{ab}|ab\>$ in the computation basis).  In other words, each vertex corresponds to a qubit initially in the state $|+\>$, and each edge corresponds to a subsequent {\sc cphase}. 

\begin{figure}[h]
    \begin{center}
    \epsfig{file=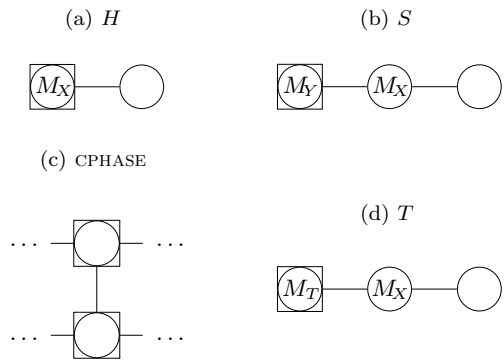} 
\vspace{0.1cm}
\caption{\label{meas-patt}
         \footnotesize{
                       Composable simulations for (a) the Hadamard gate ($H$), (b) the rotation around the $z$-axis by $\pi/2$ ($S$), (c) the {\sc cphase} and (d) the rotation around the $z$-axis by $\pi/4$ ($T$). Note that we use the {\sc cphase} to simulate itself, since it can be built in as a vertical edge of the graph. We have omitted the input classical registers and their updates for simplicity. The symbols $M_{\!X}$ and $M_{\!Y}$ indicate measurements of $\sx$ or $\sy$ on the corresponding qubits, and $M_{T}$ indicates a measurement of the observable $(\sx {\pm} \sy) / \sqrt{2}$ depending on whether there is a $\sx$ correction in the input qubit.
                       }
         }
    \end{center}
\end{figure}

As precursor to a graph-state simulation, our next step is to composably simulate a universal set of circuit elements (state preparation, measurements, and a universal set of gates), using single-qubit measurements and {\sc cphase}. 
In the circuit model, it suffices to prepare any Pauli eigenvector and measure along any Pauli basis.  Both of these can be trivially simulated in the graph-state model using single-qubit measurements.  For the universal set of gates, we choice the Clifford group generators $\{H, S\equiv e^{-i \sz \pi /4},$ {\sc cphase}$\}$ and the additional non-Clifford $T\equiv e^{-i \sz \pi /8}$. Here $\{\sx, \sz \}$ denote the standard Pauli operators. 
%
%
Figure \ref{meas-patt} shows how to composably simulate these gates, 
with the classical registers omitted for simplicity.
In Fig.\ \ref{meas-patt}, qubits are represented as circles. The boxed circles contain the quantum inputs, unboxed ones are prepared in $|+\>$, and open circles (unmeasured qubits) contain the quantum outputs.  
Edges denote {\sc cphase} gates acting on the adjoined qubits. The measurement bases for each qubit are given in the circle. 
The quantum state at the input of each pattern has known Pauli corrections labeled by the classical register $\ein$ (not shown), which depends on past measurement outcomes. In the simulation of $T$, $\ein$ is used to control one of the quantum measurements. The output quantum state also has Pauli corrections labeled by an updated string $\eout$. Each simulation pattern defines an update rule, mapping $\ein$ and measurement outcomes obtained in the pattern to $\eout$. 

\begin{figure}[h]
    \begin{center}
\epsfig{file=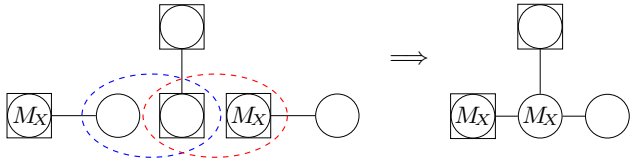}                        
\vspace{0.2cm}
\caption{\label{meas-patt:cnot}
         \footnotesize{
                       A schematic diagram of the composition of the patterns in Fig.$\,$\ref{meas-patt}(a), Fig.$\,$\ref{meas-patt}(c), and Fig.$\,$\ref{meas-patt}(a) that simulates {\sc cnot}$\,=(I\otimes H)\,${\sc cphase}$\,(I\otimes H)$. In the dashed ellipse on the left, the output of the measurement pattern simulating the first $H$ is identified with the input for the lower qubit of the {\sc cphase} simulation, whose output for the same qubit is identified with the input qubit of the simulation of the second $H$ (right ellipse). On the right is the result of the composition.
                       }
         }
    \end{center}
\end{figure}

Any circuit (sequence of gates and measurements on standard initial states) can then be simulated by composing a sequence of simulations by identifying the quantum output of one simulation (the open circle) with the input to the next (the boxed circles) and similarly for the classical registers. The combined simulation thus consists of single-qubit measurements on qubits prepared in a graph state (with the {\sc cphase} being part of the graph state preparation), giving a complete recipe for the entire graph-state simulation. Note that evolutions of single qubits and their interactions ({\sc cphase}) in the simulated circuit are represented in the graph as linear paths and the links between them, respectively. As an example, Fig.$\,$\ref{meas-patt:cnot} shows how a composition of the measurement patterns for the simulation of $H$ and {\sc cphase} leads to a new pattern for the simulation of the operation {\sc cnot}$\,=(I\otimes H)${\sc cphase}$(I\otimes H)$.

\section{Noisy graph-state computation}

We now investigate how noise at the level of the graph-state simulation maps to noise in the simulated operations and, most importantly, whether such ``simulated noise'' can be tolerated by simulating a fault-tolerant circuit described in the circuit model. We begin by mentioning a modification to the graph-state model that is necessary for fault tolerance. Since the ability to prepare fresh qubits and interact them with the existing ones is essential for all fault-tolerant constructions \cite{Aharonov96b}, instead of creating the entire graph state before computation, a minimal modification to the model is to build the required graph state dynamically as the computation proceeds \cite{Raussen01,Raussen03b,Nielsen04,Nielsen04b}. 
%
%
The simulated circuit defines a partial time ordering of the simulations and the measurements used therein, inducing a partial ordering of the qubits in the graph state.  The qubits can be added slightly before their preceding neighbors are measured, as long as the {\sc cphase} gates are applied according to the time ordering of the simulations.  
This change in the model still preserves the appealing feature of the graph-state model in that all unitary interactions are applied prior to and independent of the measurements that realize the computation.

Coming to the main part of this paper, we must analyze how physical noise affects the elementwise simulations and how the noisy simulations compose together. The elementary steps in the simulation are the preparation of $|+\>$, the {\sc cphase}, the single-qubit measurements, and the storage of qubits. Moreover, each operation belongs to a unique simulation. Thus, noise afflicting a given operation only acts within one simulation. In particular, an erroneous {\sc cphase} cannot affect two successive simulations. 

In any noise model and without loss of generality, each noisy state preparation or noisy gate can be expressed as the ideal operation followed by a \emph{noise operation}. Hence, noise operations intersperse pairs of successive ideal operations. A noise operation is a system-environment coupling, 
and it can always be 
described by some unitary joint evolution 
\be
\label{eq:expansion}
  U_{\rm fault} = I \otimes A_{0} + \sum_i P_i \otimes A_{i} \, , 
  \vspace*{-1ex}
\ee
\noindent where $P_i$ ranges over all nontrivial Pauli operators indexed by $i$ acting on the output system of the preceding ideal operation and each $A_{i}$ is an arbitrary operator acting on the environment, subject to the condition that $U_{\rm fault}$ is unitary. A noisy measurement is modeled as the ideal measurement \emph{preceded} by a noise operation given by Eq.$\,$(\ref{eq:expansion}), with $P_i$ acting on the qubits to be measured.

We first consider independent stochastic noise processes. In this case, each noise operation is {\em by assumption} acting on a separate environmental register, which is mapped to orthogonal states by the two terms in Eq.$\,$(\ref{eq:expansion}). Physically, this assumption corresponds to the requirement that a {\em record} be kept in the environment whenever faults occur, which can in principle be read to indicate the location of faults. 
In more detail, the two terms result in perfectly distinguishable environmental states, so that the corresponding states in the system {\em do not interfere} with one another, and their normalization can be interpreted as the probabilities of the first or second term in Eq.$\,$(\ref{eq:expansion}) occurring.  These two terms thus correspond to the two events of not having or having a fault.  We call the second term in Eq.$\,$(\ref{eq:expansion}) the \emph{fault operator} or simply the fault. A \emph{fault path} for the entire computation is an event occurring with some definite probability describing whether each noise operation results in a fault or not.
%

Our first goal is to show that faults within one simulation only affect that simulated operation, even though classical registers that carry the Pauli corrections and control the simulation are shared by many simulations. Consider a sequence of simulations $\{ \S_{\F_j} \}$ applied to an input $\sum_{\ein} p_{\ein} \ein \ot \P_{\ein}(\rin)$. 
Suppose some number of faults occur within $\S_{\F_1}$. 
Each term in the expansion of Eq.$\,$(\ref{eq:expansion}) of all these fault operators consists of Pauli operators acting on the simulation qubits which can be commuted to the end of the simulation (since, as shown in Fig.$\,$\ref{meas-patt}, each simulation is realized by a sequence of unitary {\sc cphase}(s) and single-qubit measurements). This results in a combined fault operator, each term in the Pauli expansion of which contains some Pauli operator acting on either the output classical registers of $\S_{\F_1}$ or its quantum output, or both.
The most general erroneous output is thus given by $\sum_{\eout} p_{\eout}^{(1)} \eout \ot \rout$ for some distribution $\{p_{\eout}^{(1)}\}$, where $\eout$ is some possibly erroneous classical output and $\rout = \E_{\eout} (\P_{\eoutid } \circ \F_1 (\rin))$, $\E_{\eout}$ is the completely positive trace non-increasing map induced by the combined fault operator on the quantum output and is conditioned on $\eout$, and $\eoutid$ labels the ideal corrections at the output in the absence of faults inside $\S_{\F_1}$ and depends on $\ein$. 
Let $\tilde{\E}_{\eout} = \P_{\eout}^\dagger \circ \E_{\eout} \circ \P_{\eoutid}$. Then, the output of $\S_{\F_1}$ can be rewritten as $\sum_{\eout} p_{\eout}^{(1)} \eout \ot \P_{\eout}(\tilde{\E}_{\eout} \circ \F_1(\rin))$. 
Hence, besides the extra $\tilde{\E}_{\eout}$, the noisy output state is of the same form as some ideal noiseless output, with the classical register reflecting the Pauli correction on the quantum state.  
In particular, this means that we can include errors in both the quantum and classical registers in $\tilde{\E}_{\eout}$ and interpret it as a simulated fault operation following the simulated $\F_1$. 
%
%
The above analysis can now be repeated to subsequent simulations, so that a simulated fault appears after each erroneous simulation. In each term labeled by $\eout$, the simulated evolution on the system and the environment is the intended computation (the sequence $\{\F_j\}$) interspersed by the action of simulated fault operators (whose particular type may depend on $\ein$ or $\eout$ at the corresponding erroneous simulation).

We pause to discuss the above argument again. The composable simulation has been described in many different ways in the literature, such as feed-forward of measurement outcomes and propagation of by-product Pauli operations. Since the classical knowledge (correct or not) of these by-product Pauli operations from one simulation step is input to the next, it is worrisome that an error in them will feed forward, inducing highly \emph{correlated} simulated faults in the simulated circuit, even if initial faults in the simulation are uncorrelated. It only takes a shift in one's perspective and inspection of the composability requirement to recognize a simpler interpretation of the error action. In particular, errors in the classical information of the by-product Pauli operations $\eout$ are {\em equivalent to} unknown Pauli errors in the quantum output $\rout$ of the erroneous simulation. The above argument takes full advantage of the equivalence and mathematically redefines $\eout$ to indicate {\em the} by-product Pauli operation, attributing any ``mismatch'' with an ideal simulation to noise acting on the quantum output of the simulation of $\F_1$ {\em alone}. {From} this point of view, the errors in classical information are localized and do not propagate. Being able to localize errors to individual simulated operations achieves a simple and direct mapping from the noise in the simulation to noise in the simulated circuit. 

We can now finish the proof of the existence of an accuracy threshold for independent stochastic noise in the graph-state model, using the threshold theorem for standard quantum computation \cite{Aharonov99comb, Knill96b,Kitaev97b,Aliferis05b}: 
In the circuit-model proof, certain fault paths are ``good'' and can be proved to give the ideal computation results. ``Bad'' fault paths form the complement of the ``good'' ones and have suppressed probability if the physical fault probability is below a certain critical value, the accuracy threshold.
Consider the noisy graph-state simulation of a \emph{fault-tolerant circuit}.
In the final output of the fault-tolerant circuit simulation, consider each $\eout$ term.
Our arguments based on composability ensures that the evolution of the quantum state is simply the intended simulated operations, interspersed by the action of faults.
Since each fault path in the simulation is mapped to a unique fault path in the simulated circuit due to error localization, good fault paths in a graph-state simulation can be defined as those resulting in good fault paths in the simulated circuit \cite{note3}. 
All other fault paths in the simulation are bad, and their probability will be suppressed below a certain accuracy threshold just as in the circuit model, because a simulated fault appears after some simulated operation only if there is at least one fault in its simulation.
Furthermore, the probability of this happening is at most the sum of the fault probabilities of all the elementary steps in the simulation.  
Then, with reference to Fig.\ \ref{meas-patt}, we note that the simulation of each gate in our universal set involves the use of one to two {\sc cphase} gates and zero to two measurements. Therefore the probability of any simulation containing faults is bounded by $p_{\rm sim} \leq 4 p$.
More specifically, if $p_0$ is the threshold value of the fault-tolerant architecture used in the circuit model and if $p \leq p_0 \, / 4$ in the simulation, then $p_{\rm sim} \leq p_0$ and 
the final measurement outcome will provide the correct computation results with the desired accuracy. 
%
This holds for each $\eout$ term in the final state of the simulation, thereby establishing a threshold lower bound of $p_0 \, / 4$ for the graph-state model. 

In the above, we have related the accuracy threshold 
in the graph-state model to that in the circuit model by the direct simulation of fault-tolerant architectures designed in the latter. However, we note that, in order to obtain the above threshold bound, we assumed that the fault-tolerant simulated circuit makes use of the same universal set as ours. In general, the same gate sets need not be used in both models, and elementary measurement patterns need to be composed to simulate a \emph{single} operation in the simulated circuit.
In particular, in most studies, {\sc cnot} rather than {\sc cphase} is used as the elementary interaction.
In this case, the measurement pattern in Fig.$\,$\ref{meas-patt:cnot} for the simulation of {\sc cnot} implies the threshold condition $p \leq p_0/5$. However, in many cases of interest this lower bound is pessimistic.
For example, in fault-tolerant designs based on self-dual CSS codes (e.g., \cite{Steane02,
Knill04, Aliferis05b}), {\sc cphase} can replace {\sc cnot} as an
alternative bitwise encoded operation and can also be used in
error correction with a small number of additional $H$
gates. Since there is no overhead for simulating single-qubit state preparation, measurement, or the {\sc cphase} in the graph-state model, the thresholds for circuits based on these codes in the circuit and graph-state models will be essentially the same. 

%
%

We now proceed to prove the existence of an accuracy threshold for the
graph-state simulation for local non-Markovian noise. We will make use
of our observation of the localization of errors
and the threshold results in the circuit model \cite{Terhal04,
Aliferis05b}. 

In the local non-Markovian error model, the noise operations still
have the form given by \eq{expansion} and they act on the system in the
same way as in the local Markovian model.  However, different noise
operations may now act on the same environmental register, and the
term acting trivially and nontrivially on the system may not map the
environmental register to orthogonal states.  Altogether, faults can
combine coherently.  
Furthermore, a fault no longer corresponds to an ``event,''
in the sense that probabilities cannot be assigned.  Instead, one imposes that the
strength of the fault operator at each location is bounded below a
certain value $\eta$---i.e. $|| \sum_i P_i \otimes A_i ||_{\rm
sup} \leq \eta$.
 
To simplify the analysis, 
we consider the \emph{purification} of the graph-state simulation, where measurements are replaced by coherent operations by attaching extra ancillary qubits.
In our noise model, noisy measurements are modeled as being ideal with noise factored into the preceding noise operations, so that changing our description of the measurements does not affect the analysis.
Likewise, the classical $2n$-bit string carrying $\ein$ can be
replaced by a $2n$-qubit register in the state $\cein$ and any
adaptive operations inside these equivalent simulations will be controlled
by this quantum register. The update of this register to obtain
$\ceout$ can also be done coherently by controlling gates from the
extra ancillary qubits and also by doing the classical processing reversibly.
We emphasize that this alternative coherent description is purely mathematical and is also employed in the circuit-model proofs in Refs.~\cite{Terhal04, Aliferis05b}.
%
%
The composable simulation $\S_\F$ is now a conjugation by a unitary
operator $S_F$ taking two inputs $\cein$ and $P_{\ein} \psiin$ and some ancillary qubits starting in the fixed state $|+\rangle^{\otimes k}$ \cite{note2}, so that $\forall \psiin$, $\forall \cein$ it acts as
\be
\begin{array}{c}
\label{eq:compdef-coh}
	S_F (\cein \otimes P_{\ein} \psiin  \otimes |+\rangle^{\otimes k} ) 
	\hspace*{15ex}\\[1.2ex]
	= \sum\limits_{i} c_{i} \, \ceout \otimes |i\> \otimes |\phi_i\> \otimes 
						P_{\eout} F \psiin \, ,
						\vspace*{-1ex}
\end{array}
\ee
\noindent where $\{ |\phi_i \> \}$ is the orthonormal basis on which measurements are to be performed, $\{ |i \rangle \}$ is the computation basis with $i$ labeling the possible measurement outcomes carried by the extra ancillas we have introduced, $c_{i}$ is the \emph{amplitude} of the $i$th term, $\ceout$ is a $2n$-qubit state that depends on $\ein$ and $i$, and $F$ is the simulated unitary operator.

Having expressed the fault-tolerant circuit to be simulated as well as
the graph-state simulation itself 
unitarily, a unitary noise operation of
the form of \eq{expansion} is inserted \emph{at every} location in the
simulation (where locations are specified by the original graph-state
simulation before the unitary idealization). The output state is a
{\em linear superposition} of states, each evolved according to a
specific set of fault operators and expanded in the eigenbasis of all measured operators (including both measurements part of the graph-state simulation and also measurements originally in the simulated circuit).  
Fault paths can again be ``good'' or ``bad,'' defined as in our discussion for independent stochastic noise. 
For each term evolved by a good fault path, a final quantum state that
will provide the correct statistics will be generated, independent of
the state of the register $|i\>$ coherently carrying the measurement information
due to the localization of errors. 
This is because, for each term in the Pauli expansion of faults acting
on $\ceout$, the register $\ceout$ can always be taken to carry the
correct Pauli correction by redefining the error acting on $P_{\eout}
F \psiin$ exactly as in our previous discussion.  Therefore, for each
term in this Pauli expansion, good fault paths in the
simulation are mapped to good fault paths in the simulated circuit that produce the ideal computation results, using the threshold theorem in the circuit model.
Hence, by linearity, the whole coherent sum of these terms will also produce the ideal computation results.

It remains to bound the {\em sup norm} of the bad fault paths of the
graph-state simulation, which can combine coherently.
Following the threshold theorem in the circuit model for local
non-Markovian noise \cite{Terhal04, Aliferis05b}, it suffices to bound
the sup norm of the ``bad'' part of a given simulation (i.e., the sum
over terms of the form $\sum_i P_i \otimes A_{i}$ in at least one location
within this simulation).
But this sup norm is simply bounded by $\eta_{\, \rm sim} \leq 4 \eta$,
where $\eta$ is a bound on the sup norm of the fault operator acting
on each location in the simulation (by the triangular inequality of the sum norm).  Thus $\eta \leq \eta_0 /4 $ is the
threshold condition for the graph-state model if $\eta_0$ is the
established threshold strength for the circuit model.  

\section{Conclusion}

To conclude, we have invoked the composability property of simulations in the graph-state model to show that faults in the graph-state simulation of any quantum circuit (and of a fault-tolerant circuit, in particular) can be viewed as affecting the simulated operations alone. Thus, the existence of an accuracy threshold for the graph-state model follows from the threshold theorem in the circuit model for the same noise process. As an aside, the same insight can be applied to other measurement-based models of quantum computation and the teleportation of gates.  
Although other proofs for the existence of an accuracy threshold in the graph-state model have already been reported for a variety of error models \cite{Raussen03b, Nielsen04b}, we believe our analysis provides an alternative, conceptually different and in many respects simpler way of thinking about fault-tolerant circuit simulations. 

We note that in optical implementations of graph-state computation \cite{Nielsen04}, gate nondeterminism and photon losses give additional sources of faults not treated in this work. 
The works in Refs.$\,$\cite{Nielsen04,Nielsen04b,Browne04} show how to control these faults by preparing 
microclusters.
%
A precise threshold analysis in this setting is pursued elsewhere \cite{Nielsen05}.


\begin{acknowledgments}
We thank Michael Nielsen and Robert Raussendorf for helpful discussions on their work in this problem. P.A. and D.L. are supported by the US NSF under grant no.$\,$EIA-0086038. D.L. is also supported by the Richard Tolman Foundation and the Croucher Foundation.
\end{acknowledgments}
\bibliography{graph-state-FT-QC-v8}

\begin{thebibliography}{27}
\bibitem[{not({\natexlab{a}})}]{note3}
\bibinfo{note}{Note that, with their locations fixed, the particular
  \emph{types} of simulated faults can be correlated since they depend on the
  classical register carrying $\ein$ and $\eout$ at each erroneous simulation,
  which is shared by many simulations. However, Ref.~\cite{Aliferis05b} shows
  that the standard threshold theorem holds for this form of noise with memory,
  which is also present in the circuit-model computation at the encoded
  levels.}
\bibitem[{not({\natexlab{b}})}]{note2}
\bibinfo{note}{These include the qubits of the graph state which are part of
  the simulation, and also extra imaginary ancillas which are used to
  coherently replace measurements by controlling the measured operators.}
\end{thebibliography}

\end{document}